\documentclass[prd,eqsecnum,showpacs,aps,nofootinbib,onecolumn,preprintnumbers,floats,superscriptaddress,floatfix,amsmath,amssymb]{revtex4-1}
\usepackage{graphicx}
\usepackage{amssymb}
\usepackage{color}
\def\physrep{phys. rep.}
\def\jcap{JCAP}
\def\mnras{MNRAS}
\def\aap{Astronomy and Astrophysics}

\begin{document}

%\title{Cosmological gravity on all scales:
%simple equations, required conditions, and a framework for model-independent modified gravity}
\title{Cosmological gravity on all scales:\\
simple equations, required conditions, and a framework for modified gravity}
\author{Daniel B. Thomas}
\affiliation{Jodrell Bank Centre for Astrophysics, School of Physics \& Astronomy, The University of Manchester, Manchester M13 9PL, UK}
\date{\today}

\begin{abstract}
The cosmological phenomenology of gravity is typically studied in two limits: relativistic perturbation theory (on large scales) and Newtonian gravity (required for smaller, non-linear, scales). Traditional approaches to model-independent modified gravity are based on perturbation theory, so do not apply on non-linear scales. Future surveys such as Euclid will produce significant data on both linear and non-linear scales, so a new approach is required to constrain model-independent modified gravity by simultaneously using all of the data from these surveys. We use the higher order equations from the post-Friedmann approach to derive a single set of ``simple 1PF'' (first post-Friedmann) equations that apply in both the small scale and large scale limits, and we examine the required conditions for there to be no intermediate regime, meaning that these simple equations are valid on all scales. We demonstrate how the simple 1PF equations derived here can be used as a model-independent framework for modified gravity that applies on \textit{all} cosmological scales, and we present an algorithm for determining which modified gravity theories are subsumed under this approach. This modified gravity framework provides a rigorous approach to phenomenological N-body simulations, and paves the way to consistently using all of the data from upcoming surveys to constrain modified gravity in a model-independent fashion. 
\end{abstract}

\maketitle

\section{Introduction}
The current cosmological paradigm, $\Lambda$CDM, contains two hypothesised forms of matter: cold dark matter, and dark energy. The existing observational evidence for these is solely gravitational in nature, so it is natural to consider whether modified gravity could be responsible for these observations. Furthermore, although General Relativity (GR) is well tested on smaller (Solar System) scales \cite{willreview}, it is less well tested on cosmological scales. Thus, there has been much interest in using cosmology to constrain alternative gravity theories \cite{1106.2476}.

One of the problems with constraining modified gravity theories is that the space of possible theories is vast. One solution to this is to use model-independent parameterisations \cite{ppf0,ppf1,ppf2,ppf3,0708.1793,0801.2431}, in order to constrain generic modifications to General Relativity. These parameterisations have been used in many works to forecast results from future surveys and to derive constraints from existing data, see \cite{1106.2476,1806.10122} for reviews.

Gravitational calculations are typically carried out in two regimes: on large scales using relativistic perturbation theory (see e.g. \cite{0809.4944}), and on small scales using the Newtonian limit and N-body simulations (see e.g \cite{bertnbody}). Nearly all of the existing parameterisations are built upon cosmological perturbation theory. Whilst this is accurate for predictions on the largest scales in the universe, it breaks down on smaller scales where the density contrast is no longer smaller than unity. The data from future surveys will include a significant contribution from non-linear scales such as Euclid \cite{1001.0061}, LSST \cite{0805.2366} and the SKA \cite{ska}, so to fully exploit these surveys for constraining modified gravity we need to be able to parameterise modified gravity on these scales.

The model-independent parameterisations can be used in several ways. The functional forms of these parameters in different theories can be used to evolve the perturbations in specific theories of gravity, and thus easily and quickly generate predictions for these theories, using common machinery (typically Boltzmann codes). In addition, it has been investigated how different functional forms or values of the parameters relate to different classes of theories, in order to constrain the maximum number of theories with the minimal amount of calculation, such as the expressions characterising the Horndeski classes of models (e.g. \cite{1404.3713}). These parameterisations also allow for generic phenomenological modifications to gravity to be investigated, and thus for modified gravity to be investigated in a maximally model-independent way. In this way, these parameterisations act as a strong null test of $\Lambda$CDM. One particular way to proceed along these lines is to bin the parameters into time and space dependent ``pixels'' that are allowed to have arbitrary values, and then perform a PCA analysis to determine which combinations are constrained by current data (e.g. \cite{1003.0001}) and future experiments (e.g. \cite{1111.3960,1501.03840}). This approach has several advantages, including acting as a potential check on systematics \cite{1003.0001}, showing where the constraining power of specific data combinations is, and requiring (almost) no assumptions as to the nature of the modifications to gravity.

The forecasts \cite{1111.3960,1501.03840} restrict themselves to linear scales, as the parameterisations are not defined on non-linear scales and there is no theoretical justification for them on non-linear scales. This significantly affects how useful these surveys will be for constraining deviations from General Relativity. Conversely, works such as \cite{1003.0001,1212.3339} (and others) use the linear theory parameterisations to justify the structure of modifications on non-linear scales; this is problematic as it isn't clear a priori what these parameters mean and whether parameterising gravity in this same way across all scales is meaningful. We will resolve this issue in this work.

The linear theory parameterisations have also been used to inspire phenomenologically modified N-body simulations \cite{1001.5184,1112.6378,1301.3255} in an attempt to calculate model-independent modified gravity predictions on non-linear scales (see \cite{0604038,0709.0307} for some earlier phenomenological modifications to N-body simulations). However, these works are lacking in theoretical justification for the modifications, in terms of when it is theoretically consistent and sufficient to parameterise the Poisson and slip equations on all scales, what the parameters mean, and whether they map to any theories of modified gravity. We will also address these issues in this work.

One existing parameterisation that is valid on non-linear scales is \cite{CliftonSanghai2018} (based on the work in \cite{1610.08039}), however it seems to only apply to theories that fit into the Parameterised Post Newtonian (PPN) framework. It isn't clear that many of the currently popular cosmological models fit into this approach, including theories with screening scales, such as $f(R)$. Another recent development is \cite{2003.05927}, which parameterises the gravitational equations based on detailed spherical collapse calculations for known screening mechanisms. This has been implemented in N-body simulations, but it seems to primarily interpolate between known theories in particular regions of theory space, preventing it from achieving sufficient generality for comprehensive null tests and definitive statements about modified gravity from future surveys.

Here we use the post-Friedmann formalism \cite{postf,thesis,Bruni:2013mua,1403.4947,longerpaper,1503.07204} to create a simple set of equations that apply on all cosmological scales and naturally incorporate both the Newtonian and perturbative limits. A key step in developing this set of equations is the absence of an intermediate scale, i.e. a scale in the universe in which neither of these limits apply, which we discuss in detail. We will use this set of equations to create a theoretically justified set of parameterised Einstein equations that are valid on all cosmological scales, including scales where the density contrast is arbitrarily large.

We recap the Newtonian limit in cosmology and the post-Friedmann formalism in section \ref{sec_recap}, before deriving our equations in section \ref{sec_main}. This is used as the basis of a framework for modified gravity in \ref{sec_mg}. In section \ref{sec_mguse} we consider some of the practicalities of this framework, including presenting an algorithm for determining whether a given modified gravity theory fits into this framework. We conclude in section \ref{sec_conc}. Note that throughout, we use ``all cosmological scales'' to mean all scales where a perturbed FLRW metric with weak fields is a reasonable description of the spacetime, and by ``non-linear scales'' we mean scales of around 10Mpc and below, where the cosmological density contrast $\delta$ becomes greater than 1.\footnote{Note that non-linear behaviour in the velocity field manifests on larger scales than in the density field.} We use ``structurally non-linear'' to refer to terms that have the form ``variable $\times$ variable'', such as $\rho \vec{v}$ or $\Phi^2$. This is to distinguish from terms such as ``$\delta$'', which can be ``non-linear'' in the sense of being arbitrarily large, but is not of structurally non-linear form.

\section{The Newtonian limit in cosmology and the post-Friedmann formalism}
\label{sec_recap}
We will take the ``Newtonian approximation'' to be shorthand for the quasi-static (i.e. not evolving quickly in time, and therefore relatively down-weighting time derivatives and terms that are relevant on horizon scales), weak field and low velocity regime; these are the physical conditions typically associated with a $\frac{1}{c}$ expansion. The Newtonian limit is the leading order equations under these approximations.

The Newtonian approximation is a good approximation for a $\Lambda$CDM universe (see e.g. \cite{1101.3555,1111.2997,Bruni:2013mua,1407.8084,1604.06065,1711.06681}), both on linear scales and on smaller scales where the density contrast can become greater than unity. This is important because Newtonian N-body simulations are a standard tool for cosmology. Recently, work has begun to investigate the Newtonian limit in cosmology more thoroughly and build towards more complete general relativistic simulations of our universe, see e.g. \cite{Bruni:2013mua,1408.3352,1604.06065,1711.06681}.

One of these approaches is the post-Friedmann formalism \cite{postf}. First proposed in \cite{postf, thesis}, the post-Friedmann formalism consists of a post-Newtonian type expansion of the Einstein equations in powers of the speed of light $c$, altered compared to a ``Solar-System'' type expansion in order to apply to a FLRW cosmology. It is thus equivalent to taking a weak-field, low-velocity, quasi-static expansion of the Einstein field equations. This approximation works well for $\Lambda$CDM cosmologies (see e.g. \cite{Bruni:2013mua,1408.3352} for calculations of the leading order corrections to this limit from N-body simulations, where these corrections are shown to be small).

The starting point for the formalism is the perturbed FLRW metric, which is considered in Poisson gauge (see \ref{sec_gauges}) and expanded up to order $c^{-5}$,
\begin{eqnarray}
 g_{00}&=&-\left[1-\frac{2U_N}{c^2}+\frac{1}{c^4}\left(2U^2_N-4U_P \right)\right]\\
g_{0i}&=&-\frac{aB^N_i}{c^3}-\frac{aB^P_i}{c^5}\\
g_{ij}&=&a^2\left[\left(1+\frac{2V_N}{c^2}+\frac{1}{c^4}\left(2V^2_N+4V_P \right) \right)\delta_{ij} +\frac{h_{ij}}{c^4}\right] \rm{.}
\end{eqnarray}
Here, the two scalar potentials have each been split into their leading order (Newtonian) ($U_N$,$V_N$) and post-Friedmann ($U_P$,$V_P$) components. The gauge freedom is chosen such that the vector potential appears in the $0i$ part of the metric, and this has also been split up into $B^N_i$ and $B^P_i$. The three-vectors $B^N_i$ and $B^P_i$ are both divergence-less, $B^N_{i,i}=0$ and $B^P_{i,i}=0$. In addition, the tensor perturbation $h_{ij}$ is transverse and trace-free, $h^i_i=h^{,i}_{ij}=0$.

As is common for post-Newtonian type expansions, the terms of order $c^{-2}$ and $c^{-3}$ are considered to be leading order and the terms of order $c^{-4}$ and $c^{-5}$ are considered to be next-to-leading order. A factor of $c^{-1}$ is attached to time derivatives. The matter content is taken to be pressure-less dust, the four-velocity of which is used to construct the energy-momentum tensor, which is also expanded in powers of $c$ (see \cite{postf} for details). The parameters describing the pressure-less dust fluid are the background density $\bar{\rho}$, the density contrast  $\delta$, and the peculiar velocity $v_i$. The assumption of pressure-less dust, applies to all of the results of this manuscript; in a universe dominated by radiation, the framework developed here will not apply\footnote{However, an effective pressure caused by shell crossing (i.e. the breakdown of single streaming) is not a problem, as this would be higher order and structurally non-linear ($\sim v^2$) \cite{1004.2488,1812.06891}, so would not contribute to the simple 1PF equations derived in section \ref{sec_main}.}. 

\subsection{Leading order equations}
The leading order ($c^{-2}$) equations from the post-Friedmann formalism correspond to the standard equations used in a Newtonian cosmology,
\begin{eqnarray}
&&\frac{d\delta}{dt}+\frac{v^i_{,i}}{a}\left( 1+\delta \right)=0\label{eq_newt1}\\
&&\frac{dv_i}{dt}+\frac{\dot{a}}{a}v_i=\frac{1}{a}U_{N,i}\label{eq_newt2}\\
&&\frac{1}{c^2a^2}\nabla^2V_N=-\frac{4\pi G}{c^2}\bar{\rho} \delta\label{eq_newt3}\\
&&\frac{2}{c^2a^2}\nabla^2\left(V_N-U_N \right)=0\label{eq_newt4}
\end{eqnarray}
Note that the time derivative in these equations is the convective derivative ($dA/dt=\partial A/\partial t+v^i A_{,i}/a$, for any quantity $A$).

A major advantage of the post-Friedmann formalism is that equations beyond those of standard Newtonian cosmology can be derived. In particular, there is also a constraint equation for the vector potential that is present at the $c^{-3}$ order,
\begin{equation}
 \frac{1}{c^3}\nabla^2B^N_i=-\frac{16\pi G \bar{\rho} a^2}{c^3}\left[(1+\delta) v_i\right]\vert_v\rm{,}
\end{equation}
where ``$\vert_v$'' denotes the pure vector, i.e. divergence-free, component of the field. This vector potential is required in the metric for the consistency of the Einstein equations in the Newtonian limit \cite{postf,1312.3638}, and it represents the leading order correction to Newtonian cosmology. It has been measured from N-body simulations \cite{Bruni:2013mua,longerpaper} and found to be small, which provides a quantitative check of the Newtonian approximation, and thus the use of Newtonian N-body simulations, for GR+$\Lambda$CDM cosmology. The smallness of this vector potential will be important later on. 

\subsection{Higher order equations}
The post-Friedmann approach can be used to construct higher order equations, beyond the set of leading order equations in the previous section. Keeping terms up to $\frac{1}{c^4}$ order results in the first post-Friedmann order (``1PF'') equations, which describe structure formation on all scales (section VIII in \cite{postf}) and include terms that are higher order in the $\frac{1}{c}$ expansion, and terms that are structurally non-linear. In addition, no terms have been removed due to any quasi-static type approximation. See \cite{postf} for full details of the relevant equations and notational conventions; here we present the equations with the background subtracted. The $00$ Einstein equation is given by 
\begin{eqnarray}
&-&\frac{1}{c^2}\frac{1}{3a^2}\nabla^2 V_N +\frac{1}{c^4}\left[\frac{\dot a}{a}\dot V_N+\left(\frac{\dot a}{a}\right)^2U_N+\frac{1}{3a^2}\nabla^2 V_N^{\;2}-\frac{5}{6a^2}V_{N,i}V_{N}^{\phantom{N},i}-\frac{2}{3a^2}\nabla^2V_P\right]\nonumber\\
&=&\frac{1}{c^2}\frac{4\pi G}{3}\bar\rho \delta +\frac{1}{c^4}\frac{4\pi G}{3}\bar\rho(1+\delta) v^2\;\text{,}
\end{eqnarray}
and the trace of the space-space Einstein equation is
\begin{eqnarray}
&&-\frac{1}{c^2}\left[\frac{2}{a^2}\nabla^2 (V_N-U_N)\right]+\frac{1}{c^4} \Bigg\{-\frac{4}{a^2}\nabla^2(V_P-U_P)-\frac{2}{a^2}U_{N,k}U_N^{\phantom{N},k}-\frac{1}{a^2}V_{N,k} V_N^{\phantom{N},k}+\frac{2}{a^2} U_{N,k}V_N^{\phantom{N},k} \nonumber\\
&&\left.+\frac{4}{a^2}V_N\nabla^2(V_N-U_N)+6\left[\frac{\dot a}{a}(\dot U_N+3\dot V_N)+2\frac{\ddot a}{a}U_N+\left(\frac{\dot a }{a}\right)^2U_N+\ddot V_N\right]\right\} 
\nonumber\\&&=-\frac{8\pi G}{c^4}\bar\rho(1+\delta) v^2\;\text{.}
\end{eqnarray}
For reference, we present the conservation equations and $0i$ Einstein equation in appendix \ref{app_1pf}.

Crucially, we note that the density contrast in these equations is not required to be small. These equations contain both the complete scalar and vector parts of linear cosmological perturbation theory, as well as containing a well-defined Newtonian limit that describes the growth of structure in a FLRW Universe on sub-horizon scales whether or not the density contrast is large. Both of these limits are derived explicitly in \cite{postf}. 

\subsection{A note on gauges}
\label{sec_gauges}
The Poisson gauge is a generalisation of the Newtonian gauge that gauge fixes the vector perturbations, so is a natural choice for the post-Friedmann approach. Moreover, as shown recently by \cite{cliftongauge}, this is one of the few gauges that can be realised without the requirement of a small density contrast. For simplicity and clarity, we present all of our results in this gauge, although we note that our modified Poisson equation (equation \ref{eq_simple1pf_pois}; see later) contains the gauge invariant density contrast. We leave to future work a full investigation of the impact of gauge choice (and conversion between gauges) on the equations presented here.

\section{Simple gravitational equations on all scales}
\label{sec_main}
Although the 1PF equations are quite complicated, there is a simpler version of them that still contains all of the required information for the two limits (perturbation theory and Newtonian cosmology). In this section we will derive these ``simple 1PF'' equations and comment on the conditions required for their validity. The resulting equations are not just a curiosity: in section \ref{sec_mg} we will use them to create a simple parameterised framework for modified gravity, such that the same parameterisation applies on all cosmological scales.

We start from the 1PF equations from \cite{postf}, and we will use the ``re-summed'' potentials as defined in that work
\begin{eqnarray}
\psi_P&=&-V_N-\frac{2}{c^2}V_P\\
\phi_P&=&-U_N-\frac{2}{c^2}U_P\\
\vec{\omega}_{P}&=&\vec{B}_N+\frac{2}{c^2}\vec{B}_P\text{.}
\end{eqnarray}
We rewrite the 1PF equations in terms of these resummed potentials (neglecting terms of order $c^{-6}$ and higher), and combine all of the ``structurally non-linear'' (see introduction) terms that are higher order in $\frac{1}{c}$ into a single schematic term, denoted $\left[\text{non-linear terms} \right]$. The time-time component of the 1PF field equations gives
\begin{equation}
\frac{1}{c^2}\frac{1}{3a^2}\nabla^2 \psi_P-\frac{1}{c^4}\left[\frac{\dot{a}}{a}\dot{\psi}_P+\left(\frac{\dot{a}}{a} \right)^2\phi_P \right ]+\frac{1}{c^4}\left[\text{non-linear terms} \right]=\frac{1}{c^2}\frac{4\pi G}{3}\bar{\rho}\delta\text{,}
\end{equation}
where the ``non-linear terms'' includes both matter and metric variables; only their non-linear structure is important for our purposes here. The time-space component of the equations can be expressed in similar fashion
\begin{equation}
\frac{1}{c^3}\left[-\frac{1}{2a^2}\nabla^2\omega_{Pi} \right ]-\frac{1}{c^3}\left[2\frac{\dot{a}}{a}\phi_{P,i}+2\dot{\psi}_{P,i} \right]+\frac{1}{c^5}\left[\text{non-linear terms} \right]=\frac{1}{c^3}8\pi Ga\rho v_i-\frac{1}{c^5}8 \pi G \bar{\rho} a \omega_{Pi}\text{.}
\end{equation}
We want to consider the scalar and pure vector parts of this equation separately, so we take the divergence to yield
\begin{equation}
-\frac{1}{c^3}\left[2\frac{\dot{a}}{a}\nabla^2\phi_{P}+2\nabla^2\dot{\psi}_{P} \right]+\frac{1}{c^5}\left[\text{non-linear terms} \right]=\frac{1}{c^3}8\pi Ga\left(\rho v_{i}\right)_{,i}\text{,}
\end{equation}
for the scalar part of the equation. We write the vector part of this equation by removing the terms that are purely a divergence, leaving
\begin{equation}
-\frac{1}{2c^3a^2}\nabla^2\omega_{Pi}+\frac{1}{c^5}\left[\text{non-linear terms} \right]\vert_v=\left[\frac{1}{c^3}8\pi Ga\rho v_i\right]\vert_v-\frac{1}{c^5}8 \pi G \bar{\rho} a \omega_{Pi}\text{.}
\end{equation}
We will also require the space-space component of the 1PF field equations
\begin{equation}
\frac{1}{c^2}\left[\frac{2}{a^2}\nabla^2\nabla^2 \left(\psi_P-\phi_P \right)\right]+\frac{1}{c^4}\left[\text{non-linear terms} \right]=0\text{.}
\end{equation}
Finally, we have the Euler and continuity equations
\begin{eqnarray}
&&\frac{dv_i}{dt}+\frac{\dot{a}}{a}v_i-\frac{\phi_{P,i}}{a}+\frac{1}{c^2}\left[\text{non-linear terms}\right]=\frac{1}{c^2}\frac{1}{a}\frac{d}{dt}\left(a\omega_{Pi} \right)\\
&&\frac{d\delta}{dt}+\frac{v^{i}_{,i}}{a}\left(1+\delta\right)-\frac{3}{c^2}\frac{d\psi_P}{dt}+\frac{1}{c^2}\left[\text{non-linear terms}\right]=0\text{.}
\end{eqnarray}

Focussing initially on the time-time and space-space equations, we now switch to Fourier space (a tilde ``$\tilde{A}$'' denotes quantities in Fourier space), in order to replace the derivatives and pull out the required structure of the scalar equations,
\begin{eqnarray}
&&-\frac{1}{c^2}\frac{1}{3a^2}k^2 \tilde{\psi}_P-\frac{1}{c^4}\left[\frac{\dot{a}}{a}\dot{\tilde{\psi}}_P+\left(\frac{\dot{a}}{a} \right)^2\tilde{\phi}_P \right ]+\frac{1}{c^4}\left[\widetilde{\text{non-linear terms}} \right]=\frac{1}{c^2}\frac{4\pi G}{3}\bar{\rho}\tilde{\delta}\\
&&\frac{1}{c^3}\left[\frac{\dot{a}}{a}\tilde{\phi}_{P}+\dot{\tilde{\psi}}_{P} \right]+\frac{1}{c^5}\left[\widetilde{\text{non-linear terms}} \right]=\frac{1}{c^3}\frac{4\pi G}{k^2}\bar{\rho}\widetilde{\left(\left[1+\delta\right] v_i\right)_{,i}}\\
&&\frac{1}{c^2}\left[\frac{2}{a^2}k^4 \left(\tilde{\psi}_P-\tilde{\phi}_P \right)\right]+\frac{1}{c^4}\left[\widetilde{\text{non-linear terms}} \right]=0\text{.}
\end{eqnarray}
We then substitute for the time derivatives in the time-time equation to arrive at two constraint equations,
\begin{eqnarray}
&&-\frac{1}{c^2}k^2 \tilde{\psi}_P+\frac{1}{c^4}\left[\text{non-linear terms} \right]=\frac{1}{c^2}4\pi Ga^2\bar{\rho}\tilde{\delta}+\frac{1}{c^4}3a^2 \frac{\dot{a}}{a}\frac{4\pi G}{k^2}\bar{\rho} \tilde{\theta}\\
&&\frac{1}{c^2}\left[\frac{2}{a^2}k^4 \left(\tilde{\psi}_P-\tilde{\phi}_P \right)\right]+\frac{1}{c^4}\left[\text{non-linear terms} \right]=0\text{.}
\end{eqnarray}
We remind the reader that the density contrast is not required to be small in these equations, hence these equations describe gravitational interactions on all scales. Note that when doing the substitution, the structurally non-linear parts of the $\widetilde{\left(\left[ 1+\delta \right]v_{i}\right)_{,i}}$ term have been moved into the ``non-linear'' terms, and we have defined $\theta=v_{i,i}$.

To proceed further, we note that the terms denoted $\left[\text{non-linear terms} \right]$ will vanish in both the small scale (Newtonian) limit, and the large scale (linear limit), because in the former case these terms are higher order in the $c^{-1}$ expansion, and in the latter case because these terms are non-linear. This leaves us with the following equations 
\begin{subequations}
\label{eq_simple1pf}
\begin{eqnarray}
&&\frac{1}{c^2}k^2 \tilde{\psi}_P=-\frac{1}{c^2}4\pi Ga^2\bar{\rho}\tilde{\delta}-\frac{1}{c^4}3a^2 \frac{\dot{a}}{a}\frac{4\pi G}{k^2}\bar{\rho} \tilde{\theta}\label{eq_simple1pf_pois}\\
&&\frac{1}{c^2}\left[\frac{2}{a^2}k^4 \left(\tilde{\psi}_P-\tilde{\phi}_P \right)\right]=0\text{,}
\end{eqnarray}
\end{subequations}
which are valid both on non-linear scales where the density contrast is large, and on larger scales where the Newtonian limit (in particular the quasi-static approximation) is not valid. The structure of these equations is familiar: they follow the same structure as the standard linear perturbation equations, except that the potentials are the re-summed potentials, and the density contrast is not required to be small. As is usual with such equations, the latter of these equations can of course be replaced with the equation $\tilde{\psi}_P=\tilde{\phi}_P$. We can combine these equations with the divergence-free part of the 1PF $0i$ equation, and the 1PF Euler and continuity equations,\footnote{In a slight abuse of notation, we will leave the continuity and Euler equations in real space, as this is the form they are typically seen in in the Newtonian limit and N-body simulations.} from all of which the structurally non-linear terms have also been dropped, yielding
\begin{eqnarray}
&&\frac{1}{2c^3a^2}k^2\tilde{\omega}_{Pi} =\left[\frac{1}{c^3}8\pi Ga\widetilde{\rho v_i}\right]\vert_v-\frac{1}{c^5}8 \pi G \bar{\rho} a \tilde{\omega}_{Pi}\\
&&\frac{dv_i}{dt}+\frac{\dot{a}}{a}v_i-\frac{\phi_{P,i}}{a}=\frac{1}{c^2}\frac{1}{a}\frac{d}{dt}\left(a\omega_{Pi} \right)\\
&&\frac{d\delta}{dt}+\frac{v^{i}_{,i}}{a}\left(1+\delta\right)-\frac{3}{c^2}\frac{d\psi_P}{dt}=0\text{.}
\end{eqnarray}
At linear order in perturbation theory, the scalars, vectors and tensors decouple from each other (e.g. \cite{kodamasas}). Moreover, the vectors and tensors can be treated as higher order as long as (as is the case in $\Lambda$CDM) no matter sources that actively generate linear vector and tensor perturbations (such as topological defects \cite{0110348}) are present. Since the vector potential term in the Euler equation is also higher order in the $\frac{1}{c}$ expansion, this suggests that this term can also be dropped in both of these two limits. This decouples the evolution of the scalar, vector and tensor components of the metric. It was shown in \cite{Bruni:2013mua,longerpaper} that the vector potential is negligibly small on all scales in $\Lambda$CDM, which justifies this choice. We now concentrate solely on the four equations containing only scalar variables, which we will refer to as the ``simple 1PF equations'',

\begin{subequations}
\label{eqn_4simple1pf}
\begin{eqnarray}
&&\frac{1}{c^2}k^2 \tilde{\psi}_P=-\frac{1}{c^2}4\pi Ga^2\bar{\rho}\tilde{\delta}-\frac{1}{c^4}3a^2 \frac{\dot{a}}{a}\frac{4\pi G}{k^2}\bar{\rho} \tilde{\theta}\\
&&\tilde{\psi}_P=\tilde{\phi}_P\\
&&\frac{dv_i}{dt}+\frac{\dot{a}}{a}v_i-\frac{\phi_{P,i}}{a}=0\\
&&\frac{d\delta}{dt}+\frac{v^{i}_{,i}}{a}\left(1+\delta\right)-\frac{3}{c^2}\frac{d\psi_P}{dt}=0\text{.}
\end{eqnarray}
\end{subequations}
These four equations give a complete (leading order) description of structure formation in both the Newtonian and perturbative limits, for any cosmology where the Newtonian limit applies and there are no linear sources of the vector and tensor in perturbation theory. If the vector potential is also small enough to not affect light bending significantly at the level of our current observations \cite{1403.4947} (which is a more stringent constraint than just being small enough for the Newtonian approximation to be good), then the scalar potential evolution described by these equations is sufficient to describe all large scale structure observables, including weak lensing, in these two limits.

However, the real power of these simple equations is that they are of much more general validity than just the large and small scale limits: In a realistic $\Lambda$CDM-like cosmology, these equations are actually valid on \textit{all} cosmological scales. This is because, if the Newtonian limit is a good description of the Universe on all non-linear scales, and the gravitational dynamics are suitably ``quasi-static'' on linear scales well within the horizon (both of which known to be true in a $\Lambda$CDM Universe) there is no ``intermediate regime'' in which the non-linear terms that we have discarded are important for structure formation. In other words, the two limits actually overlap, and thus the ``simple 1PF'' equations describe structure formation on \textit{all} cosmological scales. We will now expand on this ``intermediate regime'' and its absence in $\Lambda$CDM.

\subsection{The intermediate regime}
In this section we discuss the ``intermediate regime'' between the two limits usually used in cosmology, namely perturbation theory (large scales) and the Newtonian limit (small scales). The intermediate regime refers to any scales where both non-linearity and higher order $\frac{1}{c}$ effects are important for the gravitational equations, and thus the gravitational equations from neither limit apply.  In $\Lambda$CDM, the lack of an intermediate regime  is due to a combination of similarities between the equations in the two limits (including the decoupling of scalar, vector and tensor perturbations), the ability to ignore some terms in the perturbative equations on sub-horizon but linear scales, and the validity of the Newtonian approximation on all non-linear scales. This lack of an intermediate regime means that the simple 1PF equations apply on all cosmological scales in $\Lambda$CDM, which will also be important when using them as the basis of a parameterisation of modified gravity in section \ref{sec_mg}.

Firstly, let us consider the similarities between the two limits. Both limits are weak field\footnote{It is known that in a $\Lambda$CDM cosmology the perturbations to the metric are small \cite{1407.8084,1711.06681}, except in the vicinity of black holes and other such objects, which is a negligible fraction of the spatial volume.} regimes, and therefore the metric perturbations appear linearly in both limits. This manifests in the fact that the ``re-summed'' potentials, which appear linearly in the simple 1PF equations, describe all of the relevant metric degrees in the two limits. If the Newtonian equations (for example) were a constraint equation for $\Phi^2$ and $\Psi^2$, then there would be no simple correspondence between the two limits. For there to be no intermediate regime on linear scales, the similarity between the scalar equations needs to be even more precise. There need to be no linear terms in the Newtonian equations that aren't present in the linear perturbation equations. In addition, the terms that are present in the linear limit but not in the linearised Newtonian limit, need to be negligible below a certain scale $\left(k_*\right)^{-1}$ where perturbation theory is still valid, i.e. $\left(k_*\right)^{-1}>\left(k_{NL}\right)^{-1}$, where $k_{NL}$ is the scale on which perturbations become non-linear. In GR, these terms correspond to the time derivative of the metric potential in the continuity equation and the contribution of the velocity to the gauge-invariant density contrast. It is well known that this is the case in a $\Lambda$CDM cosmology and that the linearised Newtonian limit equations match the scalar sub-horizon perturbation equations in a dust cosmology.

Tensor-type potentials do not appear at leading order in either limit, but the vector potential is more complicated. In both limits it doesn't influence the scalar dynamics, which is another key similarity between the two limits. In principle, it is still present in the metric in the Newtonian limit and could influence observables. However, it has been shown \cite{1403.4947} that the vector potential in $\Lambda$CDM is too small to influence observables at the level of upcoming surveys; if this were not the case then the simple 1PF equations would need to be supplemented by the additional equation for the vector potential in order to be a complete description of the quantities required for cosmological large scale structure observables, such as light deflection.

That there is no intermediate regime on linear scales is implicit in the use of cosmological N-body simulations, and the known results that the output of these simulations matches that of perturbation theory on linear sub-horizon scales \cite{1101.3555}. At this stage, there is just one check left to determine the lack of an intermediate regime, which is that the Newtonian approximation holds up to the non-linear scale, i.e. that none of the terms in the 1PF equations that are higher order in the $\frac{1}{c}$ expansion contribute on scales smaller than $k_{NL}$. This has been tested using the smallness of the induced vector potential,\footnote{Note that the vector potential can be small enough for the Newtonian approximation to be valid and still large enough that it must be considered for observables. Therefore the test described above for the vector potential to be ignored for all observables \cite{1403.4947} is a stricter requirement on the size of the vector potential than that required here for the Newtonian limit to be valid.} showing that the Newtonian approximation does indeed hold \cite{Bruni:2013mua,longerpaper,1408.3352}. Thus we can see that there is no intermediate regime where the gravitational dynamics require inclusion of terms that are both structurally non-linear, and higher order in the $\frac{1}{c}$ expansion (a similar conclusion is reached differently in \cite{1711.06681}).

The importance of the smallness of the vector potential to the validity of the Newtonian approximation, and restriction of the gravitational equations to dealing with scalars only, is why the vector potential will play an important diagnostic role when determining whether a particular modified gravity theory is contained within the approach presented in section \ref{sec_mg}.

\subsection{The simple 1PF equations and N-body simulations}
The simple 1PF equations are a generalisation of the usual Newtonian cosmological equations, so it is interesting to consider what the effect might be of implementing these equations in N-body simulations in lieu of the usual Newtonian equations. Comparing the simple 1PF equations to the standard Newtonian equations, there are two key differences: the time derivative of the potential in the continuity equation, and use of the gauge invariant density contrast in the Poisson equation. The requirements during the derivation that the Newtonian limit is good, and that there is no intermediate regime, mean that implementing these equations instead of the standard N-body equations will make little difference except near the horizon scale where the $\frac{1}{c}$ expansion breaks, and therefore won't be relevant for most of the applications of N-body simulations. Since these equations reduce to the standard perturbation theory equations on large scales, then any differences between the simple 1PF and standard N-body equations on sub-horizon scales would already have been seen on linear scales in the simulation, in contrast to what is found (e.g. \cite{1101.3555}).  This result will also apply to the modified gravity parameterisation (see later). However, we expect that on scales towards the horizon, the simple 1PF equations would affect the results from N-body simulations. For example, it might be expected that the matter power spectrum in the simulation on the largest scales, which is known to be sensitive to the gauge choice, might now match that calculated in the Poisson gauge. We leave a full investigation of these issues to future work.

\section{Simple modified gravity equations on all scales}
\label{sec_mg}
The full 1PF equations govern gravitational structure formation on all scales: they don't require either limit to be specified, and they don't require the density contrast to be small. As such, these equations can be used as the basis of a framework to describe modified gravitational dynamics on all scales, for modified gravity theories where the Newtonian approximation is a good description of the small scale (non-linear) dynamics. A possible version of how these modified gravity equations might look is
\begin{subequations}
\label{eqn_full1pfmg}
\begin{eqnarray}
&&\frac{1}{c^2}\frac{1}{3a^2}\nabla^2 V_N +\frac{1}{c^4}\left[\frac{\dot a}{a}\dot V_N+\left(\frac{\dot a}{a}\right)^2U_N+\frac{1}{3a^2}\nabla^2 V_N^{\;2}-\frac{5}{6a^2}V_{N,i}V_{N}^{\phantom{N},i}-\frac{2}{3a^2}\nabla^2V_P\right]\nonumber\\
&&=\frac{1}{c^2}\frac{4\pi G}{3}A\left(a,\vec{x} \right)\bar\rho \delta +\frac{1}{c^4}\frac{4\pi G}{3}B\left(a,\vec{x} \right)\bar\rho(1+\delta) v^2\\[6pt]
&&-\frac{1}{c^2}\left[\frac{2}{a^2}\nabla^2 (V_N-U_N)\right]+\frac{1}{c^4} \Bigg\{-\frac{4}{a^2}\nabla^2(V_P-U_P)-\frac{2}{a^2}U_{N,k}U_N^{\phantom{N},k}-\frac{1}{a^2}V_{N,k} V_N^{\phantom{N},k}+\frac{2}{a^2} U_{N,k}V_N^{\phantom{N},k} +\frac{4}{a^2}V_N\nabla^2(V_N-U_N)\nonumber\\
&&\left.+6\left[\frac{\dot a}{a}(\dot U_N+3\dot V_N)+2\frac{\ddot a}{a}U_N+\left(\frac{\dot a }{a}\right)^2U_N+\ddot V_N\right]\right\} =-\frac{8\pi G}{c^4}C\left(a,\vec{x} \right)\bar\rho(1+\delta) v^2+D\left(a,\vec{x} \right)\\[6pt]
&&\frac{1}{c^3}\left[-\frac{1}{2a}\nabla^2B^N_i+2\frac{\dot a}{a} U_{N,i}+2\dot V_{N,i}\right]+\frac{1}{c^5}\bigg[-\frac{1}{2a}\nabla^2B^P_i+
4\frac{\dot a}{a} U_{P,i}+4\dot V_{P,i}+2\dot V_N U_{N,i} +4\frac{\dot a}{a} U_NU_{N,i}\nonumber\\&&+4\dot V_{N,i}V_N+\frac{1}{2a}B^N_{i\phantom{N},k}(V_N-U_N)^{,k}-\frac{1}{2a}B^{N}_{k\phantom{N},i}(U_N+V_N)^{,k}+\frac{1}{a}\nabla^2B^{N}_{i}(V_N-U_N)+\frac{1}{2a}B^N_{i}\nabla^2 V_N\nonumber\\
&&+\frac{1}{a}B^{Nk}V_{N,ki}\bigg]=\frac{8\pi G}{c^3}E\left(a,\vec{x} \right)\rho a v_i+\frac{8\pi G}{c^5}F\left(a,\vec{x} \right)\rho a\left\{v_i\left[v^2+2(U_N+V_N)\right]-B^N_i\right\}\; \text{,}
\end{eqnarray}
\end{subequations}
where $A,B,C,D,E,F$ are time and space dependent parameters that represent the effects of the modified gravity. We note that we are not presenting this as an optimal, rigorous or well-motivated parameterisation of these equations. It is merely intended to be a schematic representation of how such a set of equations might look. Considering the complexity of the equations \eqref{eqn_full1pfmg}, it is unclear how useful such a parameterisation would be. Moreover, motivated by the derivation of the simple 1PF equations in the previous section, the fact that $\Lambda$CDM seems to be a reasonable model given our observations, and the lack of an intermediate regime in $\Lambda$CDM, we note that such a parameterisation is likely to be overkill.

Instead, we will perform the same process as was performed in section \ref{sec_main} to write the parameterised gravitational 1PF equations as
\begin{eqnarray}
&&\frac{1}{c^2}k^2 \tilde{\psi}_P+\frac{1}{c^4}\left[\text{parameterised non-linear terms} \right]=-\frac{1}{c^2}4\pi Ga^2 \bar{\rho}\alpha\left(a,\vec{k} \right)\left(\tilde{\delta}+\frac{\dot{a}}{a}\frac{3}{c^2k^2}\widetilde{ \theta}\right)\\
&&\tilde{\psi}_P=\beta\left(a,\vec{k} \right)\tilde{\phi}_P+\frac{1}{c^4}\left[\text{parameterised non-linear terms} \right]\\
&&\hspace{-0.0cm}\frac{1}{2c^3a^2}k^2\tilde{\omega}_{Pi} +\frac{1}{c^5}\left[\text{parameterised non-linear terms} \right]=\left[\frac{1}{c^3}8\pi Ga\gamma\left(a,\vec{k} \right)\widetilde{\rho v_i}\right]\vert_v-\frac{1}{c^5}8 \pi G\epsilon\left(a,\vec{k} \right) \bar{\rho} a \tilde{\omega}_{Pi}\text{.}
\end{eqnarray}
Here, we have switched to Fourier space and $\alpha\left(a,\vec{k} \right)$, $\beta\left(a,\vec{k} \right)$, $\gamma\left(a,\vec{k} \right)$ and $\epsilon\left(a,\vec{k} \right)$ are time- and space-dependent functions that we don't expect to necessarily have simple functional forms in general. As before, the schematic terms representing the higher-order structurally non-linear terms are hiding the majority of the complexity in these equations, except that they now contain additional complexity due to the extra modified gravity parameters that must necessarily be present. We now neglect these terms as in section \ref{sec_main}, leaving us with 
\begin{eqnarray}
\frac{1}{c^2}k^2 \tilde{\psi}_P&=&-\frac{1}{c^2}4\pi Ga^2 \bar{\rho}\alpha\left(a,\vec{k} \right)\left(\tilde{\delta}+\frac{\dot{a}}{a}\frac{3}{c^2k^2}\widetilde{ \theta}\right)\\
\tilde{\psi}_P&=&\beta\left(a,\vec{k} \right)\tilde{\phi}_P\\
&&\hspace{-1.5cm}\frac{1}{2c^3a^2}k^2\tilde{\omega}_{Pi}=\left[\frac{1}{c^3}8\pi Ga\gamma\left(a,\vec{k} \right)\widetilde{\rho v_i}\right]\vert_v-\frac{1}{c^5}8 \pi G\epsilon\left(a,\vec{k} \right) \bar{\rho} a \tilde{\omega}_{Pi}\text{.}
\end{eqnarray}
This is equivalent to a parameterisation of the gravitational equations in the simple 1PF equations derived in section \ref{sec_main}, and should be used in conjunction with the Euler and continuity equations derived there (see equations \eqref{eqn_4simple1pf}). Before continuing we note that, as before, the scalar, vector and tensor sectors are decoupled. In particular, the only effect of the vector potential will be on photon trajectories, not on the evolution of structure in the universe. We can thus drop the vector equation as before, and work with just the scalar potentials. We comment further on this in section \ref{sec_algor}.

These parameterised equations describe the leading order gravitational dynamics in both the Newtonian and linearly perturbed limits: the equations in each limit have the same structure and contain the same parameters, therefore consistent calculations can be carried out in the two limits. The only difference is that in the linear limit, the right hand side of the parameterised Poisson equation corresponds to the gauge invariant linear density contrast, as expected in the Poisson gauge, and in the Newtonian limit the right hand side reduces to the standard Newtonian density.

More importantly, for modified gravity cosmologies where there is no intermediate regime (as is the case for $\Lambda$CDM), we can go further and apply these equations on all cosmological scales, thereby having parameterised modified gravity equations where the parameters are meaningful on all scales. This means that the same parameters can be constrained {\it using data that combines linear and non-linear scales}; this is the first time that such an approach has been shown to be valid.

For the sake of clarity, we recap the argument used to derive these equations. The starting point is the full 1PF equations, which describe structure formation on all scales and don't require the density contrast to be small, thus they are the natural choice for a parameterised set of equations to describe modified gravity on all scales. We rewrite these equations into a form that separates the terms that dominate in the large and small scale limits from the additional terms that are required in an intermediate regime, and parameterise the resulting equations. We then neglect the additional terms; crucially, the same pair of parameters describe the modified gravity effects in the two limits. Then we note that in a modified gravity cosmology with no intermediate regime, these simple parameterised equations describe the modified gravitational dynamics on all scales.

\subsection{A convenient re-writing of the parameterised equations}
Much investigation has been done into different choices of how to parameterise gravitational equations with this structure, and the relative merits of the different choices \cite{1106.2476,1806.10122}. As long as the system of equations has a constraint equation for one of the potentials, and an algebraic closure relation for the other, there are multiple choices that suffice. From here onwards we will work with the alternative parameters 

\begin{subequations}
\label{eqns_final}
\begin{eqnarray}
\frac{1}{c^2}k^2 \tilde{\phi}_P&=&-\frac{1}{c^2}4\pi Ga^2 \bar{\rho}G_\text{eff}\left(a,\vec{k} \right)\left(\tilde{\delta}+\frac{\dot{a}}{a}\frac{3}{c^2k^2}\widetilde{ \theta}\right)\\\label{eqns_final_1}
\tilde{\psi}_P&=&\eta\left(a,\vec{k} \right)\tilde{\phi}_P\text{.}\label{eqns_final_2}
\end{eqnarray}
\end{subequations}

The advantage of this choice is that it is $\phi_P$ that governs the trajectories of massive particles, as can be seen from the 1PF Euler equation above. As a result, only $G_\text{eff}$ needs to be implemented into the evolution of the N-body code; the effects of $\eta$ can all be included during post-processing steps. Therefore, these equations define the optimal way to include modified gravity parameters in N-body simulations. Indeed, some theories have zero ``slip'' (e.g. \cite{noslip}), meaning that $\eta=1$, making these theories particularly easy to simulate in this approach. Another advantage of modified gravity simulations based on this framework is that there are no equations to solve for the additional fields, and thus the N-body simulations should take a similar amount of time as $\Lambda$CDM simulations. Note that, as discussed previously, only the leading order (in terms of $\frac{1}{c}$) parts of these equations need to be implemented to model theories that have a good Newtonian limit and no intermediate regime. These two equations replace the first two equations in \eqref{eqn_4simple1pf}, and we refer to the resulting set of equations as the ``Parameterised Simple 1PF'' (PS1PF) equations.

The $G_\text{eff}$ and $\eta$ parameters will reduce to the functional forms already known for modified gravity theories on large scales, because the perturbative limit of the PS1PF equations reduces to the known linear theory parameterisation. Any complicated behaviour on small scales of additional fields that may be present is hidden in these functions, thus the time and space dependencies are not expected to reduce to simple functional forms on small scales. Since our parameterisation reduces to that in \cite{ppf2} on super-horizon scales, this parameterisation satisfies the super-horizon consistency condition \cite{bert06,0801.2431}, as long as $G_\text{eff}$ and $\eta$ tend towards finite, scale-independent functions on large scales.

For convenience (because many cosmological observables are computed in Fourier space), and to connect to the linear theory parameterisations, we have defined our modified gravity parameters in Fourier space. In principle it is always possible to convert between parameters defined in Fourier space and in real space, and indeed when there is no scale dependence the conversion will be trivial. However, in the general scale-dependent case, converting between the parameters becomes non-trivial because of the required convolutions.

An important consequence of these equations is that they justify the use of phenomenological modified gravity N-body simulations, i.e. simulations based on a parameterised Poisson equation, rather than on the equations of motion derived in a specific theory. The process used to arrive at these equations can be reversed to elucidate the conditions required for a parameterisation of the Poisson and slip equations to be a sensible and sufficient description of the gravitational dynamics; see section \ref{sec_algor}. Under the phenomenological approach, the consequences of different forms for $G_\text{eff}$ and $\eta$ can be explored using N-body simulations for these values; we discuss a few possibilities for these forms in section \ref{sec_paramforms}, including the possibility of binning $G_\text{eff}$ and $\eta$ into time and space dependent ``pixels'' that can be independently varied and constrained. Note that this does not justify the use of existing Halofit \cite{halofit} and similar $\Lambda$CDM semi-analytic fitting formulas for modified gravity cosmologies: N-body simulations still need to be employed.

\subsection{The Newtonian approximation and intermediate regime in modified gravity}
\label{sec_newtmg}
The nature of the Newtonian approximation in modified gravity theories is important if the parameterisation of the simple 1PF equations is to apply to any ``real'' theories. Moreover, the equations are at their most powerful for modified gravity theories that (like $\Lambda$CDM) have no intermediate regime. Here we examine what is known about modified gravity theories in these regards, and how it justifies the choice to use this approximation in building the parameterisation.

The starting point of the majority of modified gravity cosmologies is to assume a weak field metric as in $\Lambda$CDM, and it is usual to consider no linear-order sources of vector and tensor perturbations. In addition, most of the currently popular theories are driven by the search for dark energy \cite{1106.2476}, and assume that cold dark matter is still the dominant matter component of the universe. Cold dark matter is still non-relativistic in these modified gravity scenarios and the velocities will still be small, suggesting that the Newtonian approximation will be good in these cosmologies.

However, few detailed examinations of the Newtonian approximation in modified gravity have been performed on non-linear scales. N-body simulations have been run for several specific modified gravity theories, including f(R) and DGP (see e.g. \cite{1506.06384} and references therein) and Galileon theories \cite{1306.3219,1308.3491}. In the usual quasi-static and weak-field approach used to derive the equations used in cosmological N-body simulations, several assumptions are made. The density is taken to be the only quantity in the stress-energy tensor that contributes to the Einstein equations, and the non-scalar elements of the metric are ignored. These assumptions amount to implicitly performing a $\frac{1}{c}$ expansion, although it is usually not discussed in such terms; without this expansion, there is no reason to neglect the non-scalar parts of the metric when the density contrast is allowed to become arbitrarily large.

As a result of the implicit $\frac{1}{c}$ expansion, we note a couple of subtleties that are not normally considered, and which are relevant to the framework developed here. Firstly, the expansion needs to be performed consistently for theories with dimensionful coupling constants, which means understanding the power of $\frac{1}{c}$ that should be applied to the coupling. We will discuss this in more detail in the context of the cubic Galileon in section \ref{sec_cubic}. Secondly, for the majority of theories for which N-body simulations have been run, there remain two untested assumptions: that the Newtonian limit is valid on all non-linear scales, and that only the scalar potentials contribute to the bending of light. The exception to this is Hu-Sawicki $f(R)$ gravity \cite{husawicki}, where these assumptions have been explicitly checked \cite{1503.07204,1905.07345}. In the former of these, the post-Friedmann formalism was applied to $f(R)$ gravity, providing an alternative justification for the equations used in N-body simulations and additionally deriving the equation for the leading order correction, the vector potential. This potential was measured from $f(R)$ N-body simulations and found to be similarly negligible as in $\Lambda$CDM, showing that the Newtonian approximation is good on all non-linear scales for this theory, and that even on non-linear scales only scalar gravitational potentials are important.

The work \cite{2003.05927}, which parameterises the Poisson equation in a way that generalises the behaviour seen in N-body simulations for specific theories, is implicitly built upon the same assumptions such as the absence of relevant vector and tensor perturbations, and the lack of an intermediate regime. The derivation of the simple 1PF equations and the lack of intermediate regime described above for $\Lambda$CDM show that the choice to parameterise the Poisson equation in \cite{2003.05927} is valid for theories that fit into the framework considered here. In this sense, the parameterisation in \cite{2003.05927} is a subset of the framework considered here, and can be used to guide functional forms of the PS1PF equations (see section \ref{sec_lucas}) for theories with screening mechanisms.

One important area that has been investigated in detail is the Quasi-Static Approximation (QSA) \cite{0711.2563,0802.2999,1310.3266,0807.2449,1208.0600,1503.06831,1712.00444,1902.10687}, which corresponds to a sub-horizon approximation and treating time derivatives of the metric and extra field perturbations as small compared to their spatial derivatives \cite{1302.1193,1310.3266},
\begin{subequations}
\label{eqn_qsa}
\begin{eqnarray}
&&|\dot{X}|\leq H X\\
&&|\nabla^2 X|\gg H^2X \; \; \text{or} \; k^2 \gg H^2 \text{,}
\end{eqnarray}
\end{subequations}
where $X$ represents either the metric or field perturbations. Typically it is found to be valid on scales that are smaller than the sound horizon of the dark energy perturbations \cite{1503.06831}, but more interestingly it is found in \cite{0802.2999,1310.3266} that the QSA is best in theories that are close to $\Lambda$CDM behaviour at the background level. There have also been several investigations of the QSA in N-body simulations, see e.g. \cite{1411.6128,0905.0858,1505.03539,1506.06384}. This is an important consistency test of the simulations, and it is a partial test of the Newtonian approximation in N-body simulations because it tests some of the approximations of the $\frac{1}{c}$ expansion (corresponding to the relative down-weighting of time derivatives).

The QSA is important as it is generally equivalent to neglecting the terms in the linear perturbation equations that are not present in the Newtonian limit, therefore showing that the QSA is valid is a key part of showing that there is no intermediate regime. This connection can be understood as follows. Newtonian theory is acausal, so it can only resemble GR in a ``local inertial frame'' or ``causal region'' where the system can be considered to be slowly varying (i.e. close-to-static) and in casual contact. This is part of the physical meaning of the $\frac{1}{c}$ expansion. In GR, in a matter dominated universe, the ``local inertial frame'' is set by the Hubble horizon (and the speed of light/gravity $c$), so it is expected that in this regime the Newtonian approximation could be a good description. The Newtonian approximation is therefore naturally associated with the quasi-static conditions (equations \eqref{eqn_qsa}) applied to the metric perturbations. Put differently, the terms that are removed by the quasi-static conditions are ones that would not be expected to be present in Newtonian theory. In modified gravity theories, there is an analogous ``local inertial frame'' or ``causal region'' set by the sound speed of the dark energy perturbations, within which the QSA is valid because the system can be considered to be slowly varying on these scales \cite{1503.06831}. Therefore, within this region, the terms that can be neglected are those that would not be present in a ``Newtonian-like'' approximation to the modified gravity theory. As a result, the existence of an intermediate regime on linear scales is unlikely for theories where the QSA is valid on scales larger than $k_{NL}$. 

We can reverse the argument from \cite{1310.3266} for theories that are not close to $\Lambda$CDM behaviour. For these theories, the quasi-static approximation doesn't hold on linear sub-horizon scales, so we expect that the Newtonian approximation is less likely to be valid. In particular, this means that there is more likely to be an intermediate regime and that the vector and tensor components of the metric are likely to be sourced.

The combination of all these results suggest that the Newtonian approximation is a sensible choice for examining modified gravities that appear observationally similar to $\Lambda$CDM, and that we should expect the majority of popular modified gravity theories to fit into this framework. In section \ref{sec_algor} we make this more concrete and present an algorithm for determining whether a particular theory fits into this framework. The results discussed here for Hu-Sawicki $f(R)$ show that this theory does fit into the framework of the PS1PF equations.

\subsection{Background cosmology}
The framework developed here focusses purely on the inhomogeneities and their evolution, and assumes a $\Lambda$CDM expansion history. This choice has been made for both observational and theoretical reasons.

Although it is possible to rule out some modified gravity theories using the background expansion history, many modified gravity theories can be designed to mimic the $\Lambda$CDM expansion history (e.g. \cite{husawicki}). In addition, it has been found that the quasi-static approximation (which is an indicator of whether a theory fits into this framework) is typically better for theories with a $\Lambda$CDM-like expansion history \cite{1310.3266}.

Moreover, the background expansion is well understood theoretically and, as measurements of the expansion history become more precise, modified gravity theories that make meaningful modifications to the background will be ruled out (or $\Lambda$CDM will be). Conversely, the inhomogeneities in the Universe contain a much greater amount of information, and permit a much greater spectrum of possible behaviours that are currently much less understood. In particular, future surveys such as Euclid will deliver large data volumes on the inhomogeneities in the Universe on non-linear scales.

\section{Using and applying the modified gravity framework}
\label{sec_mguse}
The discussion so far has focussed on the derivation of the simplified 1PF equations in standard gravity, and using them as the basis of a framework to describe modified gravity. In this section we delve further into the practicalities of using the framework presented here. We present an algorithm for determining whether a given theory fits into the framework proposed here, and we apply the post-Friedmann formalism to some example theories in order to illustrate the algorithm with respect to these theories. Finally, we consider some possible functional forms for the parameters.

\subsection{Algorithm for modified gravity theories}
\label{sec_algor}
We begin by recapping the properties of $\Lambda$CDM elucidated earlier that result in there being no intermediate regime, and thus that the simple 1PF equations apply on all scales,
\begin{itemize}
\item A weak field metric is appropriate on all cosmological scales;
\item There are no sources of vector and tensor perturbations in linear perturbation theory;
\item The linearised Newtonian scalar equations have no terms that don't appear in linear perturbation theory;
\item There is a scale $\frac{1}{k_*}>\frac{1}{k_{NL}}$, below which the terms that are present in linear perturbation theory, but not present in the linearised Newtonian scalar equations, are negligible;
\item The Newtonian approximation is valid on all non-linear scales.
\end{itemize}
For any theory where all of these criteria are met, the two limits will be valid, only scalar variables need to be evolved to compute cosmological structure formation, and there will be no intermediate regime. This means that the simple 1PF equations for the scalar sector can be utilised on all scales. If in addition the following extra criterion is satisfied,
\begin{itemize}
\item The effect of the vector potential (e.g. on photon trajectories) is negligible for upcoming large scale structure observations,
\end{itemize}
then the additional constraint equation for the vector potential is not required, and the equations \eqref{eqn_4simple1pf} are a complete description of the variables required for computing large scale structure observables.

The post-Friedmann vector potential is a key diagnostic here. Not only does it check the validity of the Newtonian approximation on all non-linear scales, but it is required to be small for the scalar gravitational potentials to be the only relevant gravitational degrees of freedom for structure formation. Therefore its smallness directly relates to the non-existence of an intermediate regime. It is also the quantity that needs to be calculated for the extra criterion.

We now use these criteria, the post-Friedmann vector potential, and the work carried out for $f(R)$ in \cite{1503.07204} to build an algorithm to determine whether a given theory fits under the assumptions and approximations behind the parameterisation of the simple 1PF equations in section \ref{sec_mg}.\\

\underline{Algorithm}
\begin{enumerate}
\item Check that a weak field metric is appropriate on all cosmological scales.
\item Derive the linear perturbation equations and check that there are no sources of vector and tensor perturbations.
\item Derive the equations in the Newtonian limit using the post-Friedmann approach.
\item Check for the existence of a scale $\frac{1}{k_*}>\frac{1}{k_{NL}}$, below which the only terms that contribute significantly in the linear perturbation equations are the same terms that are present in the linearised Newtonian equations (for most theories this will be equivalent to testing the QSA).
\item Derive the equation for the post-Friedmann vector potential in the modified gravity theory, use it to calculate the vector potential from N-body simulations, and check that this is small enough for the Newtonian approximation to be valid on all non-linear scales.\footnote{To some extent, ``how small is small enough?'' is an arbitrary question. To be in line with claims about 1\% precision cosmology, we propose the conservative requirement $P_\omega(a,k)< 10^{-4}P_\Phi(a,k)$ for all $a,k$.}
\end{enumerate}

The extra criterion above for the effect of the vector potential can be tested easily from step 5,
\begin{enumerate}
\setcounter{enumi}{5}
\item Determine if the effect of the vector potential on photon trajectories is negligible for upcoming large scale structure observations; a reasonable rule of thumb for this is to compare the E-mode of cosmic shear from the vector potential to that generated by higher order deflections by the scalar \cite{0910.3786,1403.4947}
\end{enumerate}

Parts 1-4 of this algorithm can be applied without the investment of running N-body simulations, as a minimal plausibility test of whether a given theory is likely to be covered by this approach. Indeed, these steps have been checked already for many popular modified gravity models (see section \ref{sec_newtmg}). Steps 5-6 provide a more rigorous and detailed test.

Note that steps 5-6 are valuable in-and-of themselves for modified gravity theories, irrespective of determining whether a particular theory fits into the modified gravity framework described here: they test the validity of the Newtonian approximation on all non-linear scales, and test that only the scalar potentials contribute significantly to cosmological observables such as the bending of light. These are untested assumptions in the majority of modified gravity theories for which N-body simulations have been run. A further benefit of running N-body simulations for a specific theory as part of this algorithm is that they can be used to determine the full $G_\text{eff}$ and $\eta$ functions for that theory as a function of time and scale, by extracting the two sides of equations \ref{eqns_final} at different snapshots, and taking the ratio. Note that this requires outputting the gravitational potentials at each snapshot, in addition to the particle position and velocity data that is the typical output of the simulations.

In comparison to other approaches such as \cite{CliftonSanghai2018}, the full procedure to determine whether a particular theory is covered by this approach is somewhat more laborious.\footnote{For a specific model, this test should ideally be done with N-body simulations that solve the equations of motion fully (rather than which use a parameterised approach). However, once a specific theory has been tested, and it is known how to reproduce it using the parameterised approach, future simulations can be run using the parameterised approach in order to achieve the gains in simplicity and speed mentioned earlier.} However, the trade-off is that the approach here does not \textit{a priori} exclude any areas of theory space (known or otherwise), or rely on extrapolating from known theories and screening mechanisms. Instead, the restrictions in theory space are based on the phenomenology of the theories; we expect the approach here to apply to any theories that can describe a $\Lambda$CDM-like universe and are thus not already ruled observationally. As a result, a null detection of the parameters in this approach would be a stronger and more definitive constraint on modified gravity effects.

From the arguments considered here, we can envisage modified gravity theories that wouldn't be described by the parameterised simple 1PF equations. For example, a theory that generates vector or tensor perturbations at first order, perhaps due to having an additional vector\footnote{Although this probably needs to be a truly divergenceless vector field, rather than a theory like Generalised Einstein-Aether \cite{gea}, where due to the timelike constraint the vector field primarily contributes a scalar mode.} or tensor field. Alternatively, a theory such as the massive gravity theory in \cite{1202.1986} might not fit into this framework, as it was pointed out in \cite{1711.09893} that the small sound speed is likely to cause problems for the QSA.

\subsection{Post-Friedmann scalar-tensor theories and applying the algorithm}
A substantial number of the modified gravity theories under investigation are scalar-tensor theories (see e.g. \cite{1106.2476}), where an additional scalar field $\phi$ is included in the action. This includes $f(R)$ gravity, which has already been shown to fit into this framework. In this section we apply the post-Friedmann formalism to several more scalar-tensor theories, in preparation for examining them in terms of our algorithm. We will particularly be interested in the post-Friedmann equation for the vector potential. Note that throughout this section we work in the Jordan frame.

Before examining any theories in detail, we first consider what result we might expect. In the $f(R)$ case, the constraint equation for the vector potential is the same as in GR; the (inhomogeneous) additional scalar field does not enter. The vector potential still has a different value in $f(R)$ because the scalar field causes the density and velocity fields to evolve differently. This result is not surprising: if the scalar field only enters the leading order Einstein equations linearly, then no matter what derivatives are applied to it, a divergence-free quantity cannot be created, and thus the divergence-free part of the $G_{0i}$ Einstein equation will not contain the scalar field. The scalar fields in these theories typically behave like the scalar metric potentials in the presence of non-relativistic matter, so it is usual to expand them in powers of $c$ as (see e.g. \cite{1501.01985, willreview})
\begin{equation}
\label{eqn_scalarexpand}
\phi=\phi_0(t)\left(1+\frac{1}{c^2}\phi_1(\vec{x},t)+\frac{1}{c^4}\phi_2(\vec{x},t)+...\right)\text{,}
\end{equation}
where $\phi_0$ is the background cosmological value of the field. Note that due to the weak field approximation, the leading order inhomogeneous part of the scalar field is assigned order $c^{-2}$, as was done in the post-Friedmann expansion for $f(R)$ \cite{1503.07204}. Assuming an expansion of this form, the inhomogeneous scalar field can only appear linearly in the leading order Einstein equations, and thus we expect the constraint equation for the vector potential to have the same form as in GR, except for a possible re-scaling of $G$ by the background field.

There is a possible loophole in this argument, which is that the scalar field could contain a dimensionful coupling constant that is assigned a positive power of $c$ in the expansion. A theory with such a coupling can have leading order terms that are structurally non-linear in the inhomogeneous quantities, and thus contribute a divergence-free term to the $0i$ equation despite only having scalars present in the theory. We will discuss this case in more detail below when applying the post-Friedmann approach to the cubic Galileon.

\subsubsection{Post-Friedmann Brans-Dicke}
The prototypical scalar-tensor theory is Brans-Dicke theory \cite{bransdicke}, where the modified action is
\begin{equation}
S=\frac{1}{16 \pi G}\int d^4 x \sqrt{-g} \left(\phi R -\frac{w}{\phi}(\nabla\phi)^2 \right) +S_m \text{,}
\end{equation}
where $S_m$ denotes the matter action and $(\nabla\phi)^2=g^{\mu\nu}\nabla_\mu \phi\nabla_\nu\phi$. For simplicity (and because it is true for the Galileon case), we will take $w$ to be a constant, and not a function of $\phi$. This choice does not change the equation for the vector potential that we will derive. The Einstein equations and scalar field equation are given by
\begin{eqnarray}
&&\phi G_{\mu \nu}-\nabla_{\mu}\nabla_{\nu}\phi +g_{\mu \nu}\square \phi-\frac{w}{\phi}\nabla_\mu \phi \nabla_\nu \phi+\frac{w}{2\phi} g_{\mu \nu}\left(\nabla \phi \right)^2 =\frac{8\pi G}{c^4} T_{\mu \nu}\\
&&\left(3+2w \right)\square \phi=\frac{8\pi G}{c^4} T\text{.}
\end{eqnarray}
We now expand these equations according to the post-Friedmann description, see \cite{postf} for the components of the Einstein and stress-energy tensors, and we expand the scalar field as in equation \eqref{eqn_scalarexpand}. Taking terms up to order $c^{-3}$, and subtracting off the background (homogeneous) terms, the equations become
\begin{eqnarray}
&&\frac{1}{c^2}\nabla^2V_N=\frac{4\pi G}{\phi_0 c^2}a^2\bar{\rho} \delta+\frac{1}{2c^2}\nabla^2\phi_1 \\
&&\frac{1}{c^2} \nabla^2\left(V_N-U_N \right)=\frac{1}{2c^2}\nabla^2\phi_1\\
&& \frac{1}{c^3}\nabla^2B^N_i=-\frac{16\pi G \bar{\rho} a^2}{\phi_0 c^3}\left[(1+\delta) v_i\right]\vert_v\\
&&\frac{1}{c^2}\left(3+2w \right)\nabla^2 \phi_1=\frac{8\pi G}{\phi_0 c^2} \bar{\rho}\delta\text{.}
\end{eqnarray}
The structure of these equations match those obtained in PPN. As anticipated above, the constraint equation for the vector potential is the same as in GR, except for a re-scaling of G by the background field $\phi_0$.

\subsubsection{Post-Friedmann cubic Galileon}
\label{sec_cubic}
We now consider the cubic Galileon as an example of a more complicated scalar-tensor theory. Following \cite{1501.01985}, we define $Y=-\frac{1}{2}\left(\nabla \phi \right)^2$ and write the action as
\begin{equation}
S=\int d^4 x \sqrt{-g}\frac{1}{16\pi G}\left(R\phi -\frac{w}{\phi}(\nabla\phi)^2-\frac{\alpha_3}{4}\frac{Y\square\phi}{\phi^3} \right)\text{.}
\end{equation}
The Einstein and scalar field equations are give by
\begin{eqnarray}
&&\hspace{-0.3cm}\phi G_{\mu \nu}\hspace{-0.06cm}-\hspace{-0.06cm}\nabla_{\mu}\nabla_{\nu}\phi +g_{\mu \nu}\square \phi\hspace{-0.06cm}-\hspace{-0.06cm}\frac{w}{\phi}\nabla_\mu \phi \nabla_\nu \phi\hspace{-0.06cm}+\hspace{-0.06cm}\frac{w}{2\phi} g_{\mu \nu}\hspace{-0.06cm}\left(\nabla \phi \right)^2 \hspace{-0.06cm}+\hspace{-0.06cm}\frac{\alpha_3}{4}\hspace{-0.06cm}\left(\hspace{-0.06cm}g_{\mu \nu}\frac{Y\square\phi}{2\phi^3}\hspace{-0.06cm} -\hspace{-0.06cm}\frac{Y\nabla_{(\mu} \nabla_{\nu)}\phi}{\phi^3}\hspace{-0.06cm}+\hspace{-0.06cm}\frac{\square\phi \nabla_\mu \phi \nabla_\nu \phi}{2\phi^3}\hspace{-0.06cm}\right)\hspace{-0.06cm}=\hspace{-0.06cm}\frac{8\pi G}{c^4}  T_{\mu \nu}\\
&&\hspace{-0.3cm}\left(2w+3 \right)\square \phi+\frac{\alpha_3}{4\phi^2}\left(\frac{18Y^2}{\phi^2}+\frac{5\nabla_\mu\phi \nabla^\mu Y}{\phi} -\nabla^\mu \phi\nabla_\mu \square\phi- \square Y- \left(\square \phi \right)^2-\frac{Y\square \phi}{\phi}\right)=\frac{8\pi G}{c^4} T\text{,}
\end{eqnarray}
where as expected these equations differ from the Brans-Dicke equations due to the new terms involving the dimensionful coupling $\alpha_3$. We will proceed by applying the post-Friedmann approach as before, but first let us consider what the lowest order contributions to each equation are from the new terms. The leading order contributions to the different components of the Einstein equations are
\begin{subequations}
\label{eqn_leadinggrav}
\begin{eqnarray}
&&(00): \frac{\alpha_3 \dot{\phi}^2_0\nabla^2\phi_1}{16a^2c^4\phi^2_0}\\
&&(0i):\frac{\alpha_3 \dot{\phi}_0}{8c^5\phi^2_0}\left(\frac{\dot{a}}{a}\dot{\phi}_0\phi_{1,i}-\dot{\phi}_0\phi_{1,0i}-\ddot{\phi}_0\phi_{1,i}+\frac{\phi_0\phi_{1,i}\nabla^2\phi_1}{a^2} \right) \label{eqn_cubic0i}\\
&&(ij): \frac{\alpha_3 \dot{\phi}^2_0}{8c^4\phi^2_0}\left(\frac{\delta_{ij}\nabla^2 \phi_1}{2} -\phi_{1,ij}\right)\text{.}
\end{eqnarray}
\end{subequations}
The scalar field equation has many more terms at leading order; here we just note that they have order $\frac{1}{c^4}$, and schematically include terms such as $\left(\nabla^2 \phi_1 \right)^2$. See appendix \ref{app_pfcubic} for the full set of leading order post-Friedmann equations.

We now consider the issue of which power of $\frac{1}{c}$ should be assigned to $\alpha_3$. In \cite{1501.01985}, detailed consideration was given to assigning orders to $\alpha_3$; the majority of the insight developed there will apply here, so we will just note a couple of key points in order to proceed:
\begin{itemize}
\item The leading order behaviour of the metric is Newtonian, even deep inside the Vainshtein region.
\item The $\frac{1}{c}$ order of $\alpha_3$ essentially comes from specifying that the leading order terms contribute to the leading order scalar field equation.
\item The $\frac{1}{c}$ order of $\alpha_3$ is independent of whether one is dealing with the ``standard'' or ``dual'' formulation, and which Vainshtein limit is being considered. As a result, the leading order gravitational and scalar field equations in the $\frac{1}{c}$ expansion are the same in the different Vainshtein limits.\footnote{Taking the different Vainshtein limits is a simplification that allows a solution to be obtained for the scalar field. We do not require this in the present work as we wish the equations to be valid at all locations, irrespective of proximity to the Vainshtein radius.}
\end{itemize}
Following this logic for the post-Friedmann case considered here, we assign an order $c^2$ to $\alpha_3$. This agrees with the order assigned in the PPNV case, as might be expected: it would be strange if this parameter required a different order in the two expansions, because the PPNV expansion should result from taking the ``no cosmological evolution'' limit of the post-Friedmann equations (and then expanding with respect to the Vainshtein radius). Since $\alpha_3=\frac{M_p}{\Lambda^3}$ (see \cite{1501.01985}), this is the same order that would be anticipated using $M^2_p\sim\frac{1}{c^4}$,  (as appears in front of the stress-energy tensor in the Einstein field equations), which makes sense as we are not expanding in the $\Lambda$ scale. The same consideration of the power of $M_p$ comprising $\alpha$ yields the correct $\frac{1}{c}$ order for both the quartic and quintic Galileons as well. Thus the assignment of $c^2$ to $\alpha_3$ appears to be a consistent and robust choice.

Using this order for $\alpha_3$, the $\frac{1}{c^4}$ terms in the scalar equation all contribute at the same order as the Brans-Dicke terms, by design, matching the result from \cite{1501.01985}. In addition, the leading order additions to the gravitational equations elucidated in equations \eqref{eqn_leadinggrav} will also contribute to the leading order gravitational equations. This last result is different to the PPNV case, and arises due to the cosmological evolution of the background value of the field. Of course, if this evolution is slow, we would expect the standard Brans-Dicke term to dominate.

This might seem like an un-necessarily laborious process, just to select the leading order terms. However, we note that in the PPNV case, the dimension of the coupling is important because it shows that at leading order there is no contribution to the gravitational equations beyond those of Brans-Dicke theory \cite{1501.01985,ppnv2}. In that case, just taking the leading order Galileon terms (as decided by applying the weak field and quasi-static approximations with no consideration for the order of $\alpha_3$) would result in the wrong equations. Deriving the equations in the consistent manner described here allows us to be sure of whether these terms should be present at leading order. In the FLRW situation considered here, unlike the Minkowski situation that is the basis of the PPNV approach, the background scalar field is time-dependent, which is why we obtain a different result to the PPNV case. If it were constant, there would be no Galileon terms contributing to the Einstein equations beyond the usual Brans-Dicke terms. Note that there is no such issue for the case of $f(R)$ gravity, because there is no new dimensionful coupling constant in the action, only the usual $\frac{8\pi G}{c^4}$ term that is present in the GR case.

We can obtain our desired result for the vector potential by taking the pure vector part of the cubic Galileon contribution to the $0i$ equation, which will contribute to the constraint equation for the vector potential. Knowing the order of $\alpha_3$, we define $\alpha_3=c^2\alpha^*_3$ in order to make the order of the extra terms explicit. From equation \eqref{eqn_cubic0i}, the additional terms are
\begin{equation}
\frac{\alpha^*_3 \dot{\phi}_0}{8c^3\phi^2_0}\left(\frac{\dot{a}}{a}\dot{\phi}_0\phi_{1,i}-\dot{\phi}_0\phi_{1,0i}-\ddot{\phi}_0\phi_{1,i}+\frac{\phi_0\phi_{1,i}\nabla^2\phi_1}{a^2} \right)\text{.}
\end{equation}
We can see that these additional terms are of order $\frac{1}{c^3}$, so they will contribute to the leading order $0i$ equation.The first three terms are exactly as anticipated earlier, and have no divergence-free part. However, consider the fourth term: $\nabla\times\left(A\nabla B\right)$ contains $\nabla A \times \nabla B$, so the curl will not vanish. Instead we will have a term of the form $\nabla \nabla^2\phi_1 \times \nabla\phi_1$. The full vector constraint equation thus deviates from the Brans-Dicke form,\footnote{This difference to Brans-Dicke and $f(R)$ gravity can be understood in terms of the phenomenology of the theory. The dimensionful coupling constant results in a theory that clusters much more strongly.}
\begin{equation}
\label{eqn_cubicvec}
 \frac{1}{c^3}\nabla^2B^N_i=-\frac{16\pi G \bar{\rho} a^2}{\phi_0 c^3}\left[(1+\delta) v_i\right]\vert_v-\frac{\alpha^*_3 \dot{\phi}_0}{8c^3a^2\phi_0}\left[\phi_{1,i}\nabla^2\phi_1\right]\vert_v\text{.}
\end{equation}
This equation can be used to extract the vector potential from cubic Galileon N-body simulations. This quantity has several uses: firstly it is a consistency check of the approximations used in this approach and is part of the algorithm described above for determining whether the cubic Galileon fits into the modified gravity framework described here, which we will use in the next section. In addition, this vector potential should be taken into account when ray-tracing through N-body simulations \cite{1403.4947} and can generate unique lensing signals \cite{1403.4947,rotest}. We leave it to future work to perform this extraction.

We remind the reader that the vector potential equation is not required to have the Brans-Dicke (or GR) form in order for the theory to fit into the modified gravity approach described here. However, as we will describe in the next section, for theories where the equation matches that in Brans-Dicke, it is possible to construct a plausibility argument for determining whether these theories fit into the framework described here that does not require running N-body simulations. For theories where the vector equation does not match Brans-Dicke (or GR), no such shortcut can be made.

\subsubsection{Applying the algorithm}
Having applied the post-Friedmann formalism to these theories, we can now start examining them using the algorithm in section \ref{sec_algor}. We note that points 1 and 2 of the algorithm naturally come out of the extensive work on linear perturbation theory in scalar tensor theories, see e.g. \cite{1106.2476,0209140,1208.0600}.

Steps 3-4 require us to compare the equations in the perturbative limit with the linearised equations from the Newtonian limit. From the equations in the literature, or applying perturbation theory to the equations above, we can see that there are additional terms present in the perturbation equations. Even for Brans-Dicke theory, there are a substantial number of extra terms \cite{1106.2476,0209140}, and there are even more for the cubic Galileon \cite{1208.0600}. However, the weak-field description that applies in both limits strongly shapes the form of these extra terms: since the metric potentials and the inhomogeneities in the scalar field are all order one, there can be no structurally non-linear terms, and all of the additional terms will involve time derivatives, either because of pre-factors such as $\frac{\dot{a}}{a}$ or $\dot{\phi}_0$, or because they contain time derivatives of the perturbations themselves. For example, some of the additional Brans-Dicke contributions to the ${00}$ equation have forms such as \cite{1106.2476}
\begin{equation}
\frac{\dot{a}}{a}\dot{\phi}_1\;\;\text{;}\;\;\;\;\;\; \omega \phi_1\left(\frac{\dot{\phi}_0}{\phi_0}\right)^2 \text{.}
\end{equation}
These additional terms are of course exactly the terms that are neglected in the QSA, as discussed in section \ref{sec_newtmg}. This simplifies step 4 of the algorithm considerably, due to the substantial effort that has gone into examining the QSA (see section \ref{sec_newtmg} above, or e.g. \cite{1908.03430} for a recent summary): as long as the terms that are missing in the linearised Newtonian equations are those associated with the QSA (as they are in this case), then step 4 simply reduces to checking the QSA, which is known to be fine for the theories considered here (see e.g. \cite{1208.0600}). So these theories satisfy steps 1-4 of the algorithm.

We now move on to considering steps 5-6 of the algorithm. These tests involve checking the smallness of the vector potential on non-linear scales, and comparing it to the scalar quantities in order to determine the validity of the Newtonian approximation, and whether the vector potential can be ignored for the purposes of calculating observables such as cosmic shear.

For Brans-Dicke, we note that the expression for the vector potential is the same as in General Relativity, except for the change to Newton's constant from the background field. We know from e.g. \cite{1310.3266} that this background field cannot be changing quickly in time if the QSA is valid, so we expect that this time dependent change to Newton's constant will not significantly change the value of the vector potential. We know that the vector potential is small in $\Lambda$CDM, which means that the amplitude of the source term (composed of the density and velocity fields) must differ substantially from that in $\Lambda$CDM for the vector potential to be large in Brans-Dicke. These fields are typically within a factor of 2 of the $\Lambda$CDM values for observationally viable theories \cite{1506.06384}, so any Brans-Dicke theory with a vector potential much larger than in $\Lambda$CDM is unlikely to be a good fit to observations, and we can expect the vector potential to be small for observationally sensible Brans-Dicke theories.

This argument showing that Brans-Dicke theories satisfy the requirements on the vector potential is approximate, but doesn't require N-body simulations to be run. Unfortunately, because the equation for the vector potential in the cubic Galileon case is more complicated than just being sourced as in GR, it is difficult to construct a similar argument for the cubic Galileon. In this case, it is required to run N-body simulations (such as \cite{1306.3219}) and extract the vector potential using equation \eqref{eqn_cubicvec}. Carrying this out involves calculating the elements on the right hand side of equation \eqref{eqn_cubicvec} from snapshots that represent the N-body simulation at a particular moment in time. In GR and $f(R)$ this just requires extracting the GR term, and this has been done by using tesselations \cite{dtfe} to extract the momentum field $\rho \vec{v}$ \cite{Bruni:2013mua,longerpaper,1503.07204}, from which the pure vector part can be calculated. In practice, this can be done to reasonable accuracy using standard CIC \cite{cic,1503.07204} type methods. The power spectrum of the vector potential can then be compared to that of the scalar potentials. The additional terms in equation 5.14 will require additional quantities to be outputted from the N-body simulation, in particular $\vec{\nabla} \phi \nabla^2\phi$. Terms like this require more care to be calculated because, due to the window function, the term ideally needs to be calculated on a grid as a single quantity (rather than extracting the two fields separately and multiplying them together). 

If the vector potential calculated from the N-body simulations is found to be large, then the premise on which these simulations are run needs to be investigated more carefully, so this check should be done irrespective of determining whether the cubic Galileon fits into the modified gravity framework described here. However, we expect that steps 1-4, and the similarity of the density and velocity fields to those in $\Lambda$CDM, mean that the vector potential will be found to be small enough to justify the Newtonian approximation.

Although steps 1-4 look very promising, we cannot be sure that the cubic Galileon fits into this approach until this final test in step 5 has been done. The advantage of performing the test for step 5 is that we get the test in step 6 at the same time. Whilst it looks plausible that the vector potential will be found to be small enough to justify the Newtonian approximation, it could still be large enough to contribute to observables, and thus provide an observational discriminant between the cubic Galileon and $\Lambda$CDM \cite{1403.4947}.

\subsection{Different functional forms for the parameters}
\label{sec_paramforms}
Whilst on linear scales the parameters $G_\text{eff}$ and $\eta$ can be analytically calculated for specific theories or classes of models, on non-linear scales they are best thought of as emergent prescriptions for the complicated non-linear dynamics within particular theories. Here we mention a few sensible choices for the functional forms of the parameters in the simple 1PF equations, including some known cases and some possible functional dependencies that are worth investigating.

\subsubsection{Linear scales}
In the linear perturbation limit, the parameters in the simple 1PF equations will reduce to the known functional forms in linear theory. 

\subsubsection{Inclusion of specific scales}
\label{sec_lucas}
In \cite{2003.05927}, building on earlier work \cite{1608.00522}, the authors propose specific functional forms for the parameterised Poisson equation, and show that specific choices for some of the parameters can recreate the phenomenology of particular theories that have been studied. These particular forms are based around detailed studies of spherical collapse in theories with screening mechanisms. The modifications to the Poisson equation take the form
\begin{equation}
b\frac{k}{k_0}^a\left(\left[1+\left(\frac{k_0}{k}\right)^a \right]^{1/b} -1\right)\text{,}
\end{equation}
where $a$ and $b$ are time varying constants, and $k_0$ is a particular scale associated to the modifications (which is also expected to be time dependent).

An alternative approach was suggested in \cite{ppf0}, where the authors use $\frac{k^3}{2\pi^2}P(a,k)=\Delta(a,k)$ to measure the degree of non-linearity at a particular scale. This defines the scale of non-linearity $k_{NL}(a)$ through $\Delta(a,k_{NL})=1$. The validity of this form for predicting the power spectrum has been tested (e.g. \cite{0902.0618}; see \cite{1608.00522} for a nice summary) and found to be good for weakly non-linear scales. It would be interesting to investigate functional forms for the modifications to the Poisson equation as a function of $k_{NL}(a)$: these functions would naturally incorporate time dependence and return to the linear forms on the appropriate scales.

\subsubsection{Environmental dependence}
Rather than explicitly including specific scales, theories with screening mechanisms can be understood by allowing the modifications to the Poisson equation to have additional environmental dependence. This was used in \cite{0903.1292} and tested in \cite{0905.0858} in the form
\begin{equation}
\nabla^2 \Psi=4\pi G a^2 \bar{\rho}\delta\left(1+\Delta(\delta) \right)\text{,}
\end{equation}
where the modification $\Delta(\delta)$ is explicitly dependent on the local density, which naturally occurs in the Vainshtein screening mechanism. This form of $\Delta(\delta)$ was constructed to represent DGP, and included the ``crossover scale'' $r_c$ explicitly, and an approximate form was found to be given by $\Delta(\delta)=\frac{2}{3\beta(t)}\left(\sqrt{1+0.3\delta} -1\right)$ (where $\beta(t)$ is a time dependent function of the background cosmology).

Another form of environmental dependence that would be interesting to explore is a dependence on the local acceleration, which links to MONDian \cite{mond1983} theories, although an examination of how the post-Friedmann approach applies to such theories is beyond the scope of this paper.

\subsubsection{Maximally phenomological pixels}
As described in the introduction, a standard approach to exploring constraints on parameters whose functional forms are not known, in a maximally model independent fashion, is to express the parameters as piecewise constant functions in a set of bins, or pixels, in time and space. This technique has been successfully used for modified gravity, \cite{1003.0001,1111.3960,1501.03840} and has also been used to constrain the dark matter equation of state in a model-independent fashion \cite{1802.09541}.

As well as being the most model-independent way to analyse observational data, pixelised functions can be useful to understand the phenomenology of different theories, by isolating the ranges in scale and time where changes to the gravitational law result in specific effects. Similarly, when used in combination with a specific dataset (or forecasted dataset), a pixelised approach can determine at which times and scales that dataset is most sensitive to the modifications.

\subsubsection{Hybrid pixels}
Since the functional forms of the parameters are well known on linear scales, a hybrid scheme can be defined, where the pixel boundaries are defined with respect to $k_{NL}(a)$ rather than $k$. This reduces the number of pixels required by making the time dependence implicit, and naturally returning to known forms on linear scales. Schematically, such hybrid pixels would have the form
\begin{eqnarray}
&&G_\text{eff}=G_\text{eff,linear} \; \text{;} \;\;\; k< A_0 k_{NL}(a)\\
&&G_\text{eff}=G_\text{eff,1}  \; \text{;} \;\;\;  A_0 k_{NL}(a)< k< A_1 k_{NL}(a)\\
&&G_\text{eff}=G_\text{eff,2}  \; \text{;} \;\;\;  A_1 k_{NL}(a)< k< A_2 k_{NL}(a)\\
&&G_\text{eff}=G_\text{eff,3}  \; \text{;} \;\;\;  A_2 k_{NL}(a)< k\text{,}
\end{eqnarray}
for a set of constants $\{A_0,A_1,A_2,G_\text{eff,1},G_\text{eff,2},G_\text{eff,3}\}$ with no time dependence. It would be interesting to see whether this approach more easily captures the phenomenology of existing theories.

\subsubsection{Dark energy models}
One approach to generalising modified gravity is to examine scalar field dark energy models in increasingly general or phenomenological ways, which culminates in the Effective Field Theory of Dark Energy approach (EFTofDE; see e.g. \cite{1404.2684} for a review). In this approach, the distinction between dark energy and modified gravity essentially vanishes, meaning that similar considerations are relevant for these models and the approach taken here. In particular, for any scalar field where a weak field expansion is appropriate, then in the two limits one will naturally arrive at a Poisson equation with extra terms, as discussed above. As a result, as these scalar field models are included in simulations with increasing generality (see e.g \cite{1012.0002,1811.01519,1910.01104}), these simulations typically incorporate a phenomenologically modified Poisson equation, therefore these simulations will be similar to an implementation of the framework discussed here.

In principle, these phenomenological Poisson equations could be used to inform the choice of functional form for $G_\text{eff}$, however care should be taken for several reasons. Firstly, we note that in many of the dark energy scalar field models that are studied with these codes, the background evolution must also be modified and is usually of similar or greater importance than the modified Poisson equation. Secondly, to date these works typically include some form of ``linear extrapolation'' at some point, where non-linear terms in the equations are neglected, or linear theory results are used to inform the form of the modifications or parameters. Of course, one could trivially extend the framework in this paper to include an arbitrary phenomenological expansion history, but there is no model-independent way to relate the expansion history to $G_\text{eff}$ and $\eta$, so we prefer to deal solely with the inhomogeneities for the reasons discussed earlier.\footnote{In essence this highlights the difference between a phenomenological code being used for a model-independent exploration, and being used as a computationally cheap and easy way to run simulations for specific models.}

Due to the similarities, we expect that in any of these scalar field models where a weak field expansion is appropriate, the Newtonian approximation is good, and there is no intermediate regime, then one will naturally arrive at equations for the inhomogeneities that fit into the framework developed in this work. It would be interesting to study these dark energy models in detail with the formalism and philosophy used in this paper; in particular, it may be possible to construct a ``small scale EFTofDE'' using a post-Friedmann type expansion.

\section{Discussion}
\label{sec_conc}
We have presented a simplified set of equations that describe structure formation on all cosmological scales and that do not require the density contrast to be small, equations \ref{eq_simple1pf}. This set of equations depends on the validity of the Newtonian approximation in a given cosmology, and requires that there is no ``intermediate regime'' (where both the Newtonian limit and perturbation theory fail). This is the case for a matter dominated $\Lambda$CDM cosmology.

We have used these equations to create a simple framework for modified gravity (equations \ref{eqns_final}) that is valid on \textit{all} cosmological scales, and which is known to include currently popular models such as $f(R)$. This set of equations is important for upcoming cosmological surveys such as Euclid \cite{1001.0061}, as it allows them to constrain model-independent modified gravity by simultaneously using the data from all scales, not just linear scales. This maximises the constraints on model-independent modified gravity that are achievable by these surveys.

In section \ref{sec_algor} we present an algorithm for investigating the validity of our approximations for a given modified gravity theory, in particular whether there is an intermediate regime. This algorithm determines whether a theory of gravity is subsumed under this approach, and therefore whether constraints determined using this parameterised approach are applicable to that theory of gravity. We have illustrated this algorithm by applying the post-Friedmann approach to the cubic Galileon, and calculating the equation for the vector potential that is required for the algorithm. We note that throughout this work, we are primarily focussed on modified gravity theories that replace dark energy, and thus on cosmologies that still contain dark matter. We will investigate the application of the post-Friedmann formalism to modified gravity theories without dark matter in future work.

An important corollary of the equations and framework developed here is that it is theoretically consistent to perform phenomenological N-body simulations based on a modified Poisson equation, including work that was performed (but not justified) in \cite{1001.5184,1112.6378,1301.3255}. In particular, the derivation outlined here (and used to construct the algorithm in section \ref{sec_algor}) elucidates the conditions required for a parameterisation of the Poisson and slip equations to be a sensible and sufficient description of the gravitational dynamics. 
 
 We have presented some functional forms for the parameters that could be used in simulations, including the possibility of exploring the modified gravity parameter space by binning these parameters into time and space dependent ``pixels'', and examining the resulting phenomenology. This pixel approach enables forecasts (and data analysis) to understand which scales and redshifts different surveys (and types of surveys) are sensitive to.

The set of equations in this work is the starting point for bringing the full constraining power of future surveys to bear on modified gravity theories in a model-independent way. This will provide a significant null test of the $\Lambda$CDM paradigm and enable general conclusions about gravity to be drawn from the data, rather than statements about specific models or areas of parameter space.

\section*{Acknowledgements}
The author thanks Ali Mozaffari, Johannes Noller and Michael Kopp for useful comments on the manuscript, and Richard Battye, Francesco Pace, Sankarshana Srinivasan and Marco Bruni for useful discussions. The author acknowledges support from Science and Technology Facilities Council (STFC) grant ST/P000649/1.

\appendix
\section{1PF equations}
\label{app_1pf}
For brevity in section \ref{sec_recap} we present a subset of the 1PF equations. Here we present the additional equations that are required for section \ref{sec_main}.\\
1PF Continuity equation,
\begin{eqnarray}
\frac{d\delta}{dt} +\frac{v^{i}_{\;,i}}{a}(\delta+1)+\frac{1}{c^2}\left[(\delta+1)\left(\frac{1}{a}v^jU_{N,j}-\frac{\dot a}{a}v^2+3\frac{dV_N}{dt}\right)\right]=0\;.
\end{eqnarray}

1PF Euler equation,
\begin{eqnarray}
&&\frac{dv_i}{dt}+\frac{\dot a}{a}v_i-\frac{U_{N,i}}{a}+\frac{1}{c^2}\bigg[v_i\frac{d}{dt}(U_N+2V_N)+\frac{2}{a}U_{N,i}(U_N+V_N)-\frac{1}{a}v^2V_{N,i}-\frac{2}{a}U_{P,i}+\frac{1}{a}v_iv^jU_{N,j}-\frac{\dot a}{a}v^2v_i\nonumber\\
&&-\frac{1}{a}\frac{d}{dt}(a B^N _i)+\frac{B^N_{j,i}v^j }{a}\bigg]=0\;.
\end{eqnarray}

1PF $0i$ Einstein equation,
\begin{eqnarray}
&&\frac{1}{c^3}\left[-\frac{1}{2a}\nabla^2B^N_i+2\frac{\dot a}{a} U_{N,i}+2\dot V_{N,i}\right]+\frac{1}{c^5}\bigg[-\frac{1}{2a}\nabla^2B^P_i+
4\frac{\dot a}{a} U_{P,i}+4\dot V_{P,i}+2\dot V_N U_{N,i} +4\frac{\dot a}{a} U_NU_{N,i}\nonumber\\&&+4\dot V_{N,i}V_N+\frac{1}{2a}B^N_{i\phantom{N},k}(V_N-U_N)^{,k}-\frac{1}{2a}B^{N}_{k\phantom{N},i}(U_N+V_N)^{,k}+\frac{1}{a}\nabla^2B^{N}_{i}(V_N-U_N)+\frac{1}{2a}B^N_{i}\nabla^2 V_N\nonumber\\
&&+\frac{1}{a}B^{Nk}V_{N,ki}\bigg]=\frac{8\pi G}{c^3}\rho a v_i+\frac{8\pi G}{c^5}\rho a\left\{v_i\left[v^2+2(U_N+V_N)]-B^N_i\right\}\;.\right.
\end{eqnarray}

\section{The leading order post-Friedmann cubic Galileon equations}
\label{app_pfcubic}
For posterity, we present here the complete leading order gravitational and scalar field equations for the cubic Galileon. According to the post-Friedmann approach, these are the equations that should be implemented in N-body simulations in order for the non-relativistic truncation of the equations to be consistent.
\begin{eqnarray}
&&\frac{1}{c^2}\nabla^2V_N+ \frac{\alpha^*_3 \dot{\phi}^2_0\nabla^2\phi_1}{16c^2\phi^2_0}=\frac{4\pi G}{\phi_0 c^2}a^2\bar{\rho} \delta+\frac{1}{2c^2}\nabla^2\phi_1 \\
&&\frac{1}{c^2} \nabla^2\left(V_N-U_N \right)+\frac{\alpha^*_3 \dot{\phi}^2_0}{16c^2\phi^2_0}\nabla^2 \phi_1=\frac{1}{2c^2}\nabla^2\phi_1\\
&&  \frac{1}{c^3}\nabla^2B^N_i=-\frac{16\pi G \bar{\rho} a^2}{\phi_0 c^3}\left[(1+\delta) v_i\right]\vert_v-\frac{\alpha^*_3 \dot{\phi}_0}{8c^3a^2\phi_0}\left[\phi_{1,i}\nabla^2\phi_1\right]\vert_v\\
&&\frac{1}{c^2}\left(3+2w \right)\nabla^2 \phi_1+\frac{\alpha^*_3}{4\phi_0^3}\left[ \frac{\dot{\phi}^2_0}{2a^2c^2}\nabla^2\phi_1 +\frac{\phi^2_0}{c^2a^4}\left(\phi_{1,ki}\phi_{1,ik} \right)- \frac{1}{c^2a^4}\phi^2_0(\nabla^2\phi_1)^2+\frac{2}{c^2a^2}\phi_0\ddot{\phi}_0\nabla^2\phi_1\right]=\frac{8\pi G}{\phi_0 c^2} \bar{\rho}\delta
\text{.}
\end{eqnarray}
We note these equations may not be consistent with those used in \cite{1306.3219}. We leave a detailed comparison between the two approximation schemes, and resulting sets of equations, to future work.

\bibliography{nonlinear_ppf}
\end{document}